\documentclass[useAMS,usegraphicx]{mn2e}
\usepackage{amssymb}
\usepackage{graphicx}
\usepackage{times}
\usepackage{amssymb}
\usepackage{epstopdf}
\usepackage{graphicx}
\usepackage{footnote}
\usepackage{threeparttable}
\usepackage{float}
\usepackage{pdflscape}

\title[Near Infrared studies of nova V5668 Sgr ]{Near Infrared studies  of the carbon-monoxide and dust forming nova V5668 Sgr }

\author[Banerjee et. al.]{D. P. K. Banerjee\thanks{E-mail: orion@prl.res.in}, Mudit K. Srivastava, N. M. Ashok ,  and V. Venkataraman  \\
Astronomy and Astrophysics Division, Physical Research Laboratory, Ahmedabad, Gujarat, India. 380009 \\}

\begin{document}

\date{Accepted YYYY Month DD.  Received YYYY Month DD; in original form YYYY Month DD}

\maketitle

\label{firstpage}

\begin{abstract}
We present near-infrared (NIR) observations of Nova V5668 Sgr, discovered in outburst on  2015 March 15.634 UT,  between 2d  to 107d after outburst. NIR spectral features are used to classify it as a FeII class of nova.  The spectra follow the evolution of the spectral lines from a   P Cygni stage to a pure emission phase where the shape of the profiles suggests the presence of a bipolar flow. A notable feature is the presence of  carbon monoxide first overtone bands which are seen in emission. The CO emission is modeled to make estimates of the mass, temperature and column density to be (0.5--2.0)$\times$ 10$^{-8}$ M$_\odot$, 4000 $\pm$ 300K and (0.36--1.94)$\times$ 10$^{19}$ cm$^{-2}$ respectively. The  $^{12}$C/$^{13}$C ratio is estimated to be  $\sim$ 1.5. V5668 Sgr was a strong dust producer exhibiting the classical  deep dip in its optical light curve during dust formation. Analysis of the dust SED yields a dust mass of 2.7 $\times$ 10${^{\rm -7}}$  $M_\odot $, a blackbody angular diameter of the dust shell of 42 mas and a distance estimate to the nova of 1.54 kpc which agrees with  estimates made from MMRD relations.


\end{abstract}

\begin{keywords}
infrared: spectra - line : identification - stars : novae, cataclysmic variables - stars : individual
Nova  Sagittarii 2015 No. 2, V5668 Sgr  - techniques : spectroscopic, photometric.
\end{keywords}

\section{Introduction}
Nova Sagittarii 2015 No. 2, equivalently  PNV J18365700-2855420 or  V5668 Sgr, was discovered by      John Seach, Australia at a magnitude of  6.0   on 2015 March 15.634 UT (Seach 2015). Spectroscopic confirmation that the transient was a classical nova of the FeII class was obtained  by Williams et al (2015)  and     Powles (Seach 2015) on March 16.27 and 16. 16.628 UT respectively. The nova's light curve (Fig. 1) showed a 6 day rise to a maximum which was reached  on March 21.6742  at V = 4.32 making it one of the brightest novae in recent times.  The initial maximum was followed by a series of dips in brightness followed by rebrightenings or secondary maxima. The quasi-periodic behavior was terminated with  the onset of a  deep dust formation phase commencing from   June  which saw a striking drop of $\sim$ 7 magnitudes in V band. At the moment of writing the nova is recovering in brightness as it emerges from the dust phase. The lightcurve is thus unique as it exhibits the combined behavior of both the J (jitter) and D (dust dip) classes of light curves (Strope et al., 2010).

V5668 Sgr has attracted considerable attention. Apart from optical spectroscopic (Williams et al. 2015) and polarimetric (Muneer et. al. 2015) studies it has been keenly followed in the NIR (Banerjee et al. 2015a,b, Walter 2015) leading to, among other results, the rather rare detection of carbon monoxide in its spectra (Banerjee et al. 2015c). In the mid-IR 5 to 35 $\mu$m range,  spectra during the early  dust nucleation phase have been obtained by SOFIA,   the NASA Stratospheric Observatory for Infrared Astronomy (Gehrz et al. 2015a). Further observations with SOFIA are continuing or planned. It has also been detected in gamma rays (Cheung et al. 2015) and in X-rays (Page et al. 2015) and has shown the peculiarity of the simultaneous existence of dust and X-ray emission (dust is expected to be destroyed by harsh UV/X-ray emission). We discuss and correlate these observed properties with our NIR data. The object is fading  slowly and should remain observable for quite some time. The present study should thus be a timely documentation of  the object's early NIR evolution.

\section{Observations}
\label{sec_observations}
We present NIR multi-epoch spectro-photometry  of V5668 Sgr between 2015 March to June commencing from 2d after discovery.   Observations  beyond June were not possible  because of the ingress of the monsoon season;  further observations will commence in October when the monsoon has fully receded.  NIR spectroscopy was done  in the 0.85 to 2.4 $\mu$m region at resolution $R$ $\sim$ 1000  with the 1.2m telescope of the Mount Abu Infrared Observatory using the Near-Infrared Camera/Spectrograph (NICS) equipped with a 1024x1024 HgCdTe Hawaii array. The NIR photometry was done using a five point dither. Since the observational  and data reduction procedures   have been described in detail in several  earlier works, we do not elaborate on it here (see Banerjee and Ashok 2002,  Banerjee et. al. 2014; Joshi et. al. 2015).  The standard stars used for these observations were  SAO 187239 (spectral type B8III) for spectroscopy and SAO 186704  (Spectral type A3 III) for photometry.  Both  these standards have  very similar sky coordinates as  the nova and were hence observed at similar airmasses as the nova. Since V5668  Sgr was  bright enough to saturate the detector, photometry was done with  the petal covers of the primary mirror partially closed  whenever it was warranted.  The log of the observations and the NIR photometry is presented in Table 1.

\section{Extinction, absolute magnitude  and distance}
From the lightcurve of Fig. 1, if we neglect the quasi - periodic oscillations and consider only the average trend of decline and extend this trend  into the dust dip region, we estimate the value of $t_{2}$ to be 100 $\pm$ 10d. The MMRD relations of della Valle and Livio (1995)and Downes and Duerbeck (2000) then give absolute magnitudes of Mv = -6.91 $\pm$ 0.4 and -6.65 $\pm$ 1.82 respectively and corresponding distances of  1.31 -- 1.76 kpc and  0.68 -- 3.6 kpc respectively.  We adopt a value of $d$ = 2 kpc in subsequent calculations.  In the above estimates  we have assumed negligible extinction ($E(B-V)$ $\sim$ 0 )  based on the $B-V$ color of  0.27 at maximum (B = 4.59, V = 4.32, AAVSO data (American Association of Variable Star Observers))  which is very close to the expected intrinsic $(B-V)$ color of 0.23 $\pm$ 0.06 expected at maximum for novae (Van der Bergh and Younger 1987). A low reddening of  $E(B-V)$ = 0.20  is also found by   Schlafly and Finkbeiner (2011)   along the entire line of sight to the nova.


\begin{figure}
\centering
\includegraphics[bb=3 88 300 380,width=3.0in,height=3.0in,clip]{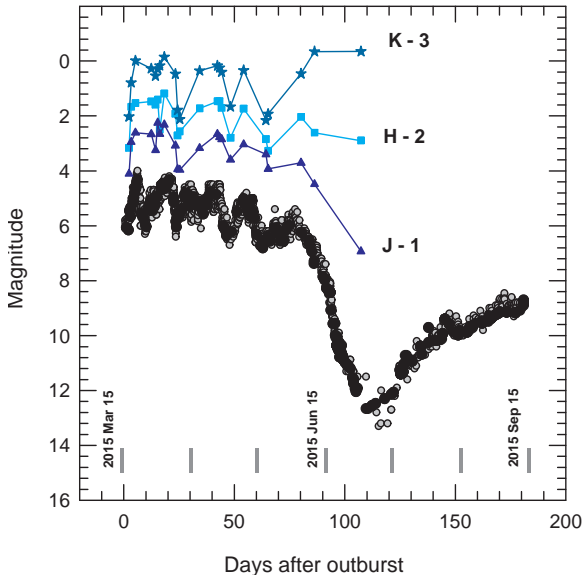}
\caption[]{ The $V$ band light curve of V5668 Sgr from  AAVSO data in black while visual estimates are  shown in gray. The $JHK$ light curves from Mt Abu data are also shown in  blue, cyan and dark blue.}
\label{fig_Lightcurves}
\end{figure}

\begin{table}
\centering
\caption{Log of the observations$^a$ from Mount Abu }
\begin{tabular}{lllllll}
\hline
Date      & Days     & Spec &     & Magnitudes &         \\
(2015)          &past    &   &$J$ &  $H$       &    $K$  \\
(UT)         & out- &    &   &            &         \\
             & burst&&&&\\
\hline
\hline

Mar 17.97	&	2.34	&\checkmark     &	5.11$\pm$0.02	&	5.17$\pm$0.14	&	5.02$\pm$0.05	\\
Mar 18.96	&	3.33	&\checkmark	&	3.95$\pm$0.15	&	3.67$\pm$0.17	&	3.79$\pm$0.02	\\
Mar 19.96	&	4.33	&\checkmark	&	----    	&	----	        &	----	        \\
Mar 20.95	&	5.32	&		&	3.61$\pm$0.13	&	3.54$\pm$0.15	&	2.99$\pm$0.01	\\
Mar 21.95	&	6.32	&\checkmark	&	----            &	----    	&	----    	\\
Mar 25.95	&       10.32	&\checkmark	&	----    	&	----    	&	----    	\\
Mar 28.03	&	12.40	&		&	3.66$\pm$0.06	&	3.48$\pm$0.06	&	3.28$\pm$0.03	\\
Mar 29.03	&	13.40	&\checkmark	&	----    	&	----    	&	----    	\\
Mar 30.00	&	14.37	&\checkmark	&	4.24$\pm$0.02	&	3.60$\pm$0.01	&	3.54$\pm$0.01	\\
Mar 30.96	&	15.33	&\checkmark	&	3.25$\pm$0.02	&	3.41$\pm$0.09	&	3.29$\pm$0.05	\\
Mar 31.94	&	16.31	&\checkmark	&	3.66$\pm$0.16	&	4.50$\pm$0.01	&	3.17$\pm$0.19	\\
Apr 1.96	&	17.29	&\checkmark	&	----    	&	----    	&	----    	\\
Apr 2.96	&	18.29	&		&	3.32$\pm$0.04	&	3.19$\pm$0.02	&	2.85$\pm$0.04	\\
Apr 6.96	&	22.29	&\checkmark	&	----    	&	----    	&	----    	\\
Apr 7.97	&	23.30	&\checkmark	&	4.08$\pm$0.02	&	3.92$\pm$0.02	&	3.47$\pm$0.05	\\
Apr 8.95	&	24.28	&		&	4.94$\pm$0.03	&	4.71$\pm$0.03	&	4.78$\pm$0.03	\\
Apr 9.96	&	25.29	&		&	4.96$\pm$0.05	&	4.57$\pm$0.26	&	5.12$\pm$0.10	\\
Apr 17.94	&	33.27	&\checkmark	&	----    	&	----    	&	----    	\\
Apr 18.94	&	34.27	&		&	4.17$\pm$0.08	&	3.73$\pm$0.29	&	3.36$\pm$0.09	\\
Apr 27.03	&	42.36	&\checkmark	&	3.66$\pm$0.05	&	3.47$\pm$0.02	&	3.17$\pm$0.06	\\
Apr 27.97	&	43.30	&		&	3.77$\pm$0.06	&	3.47$\pm$0.01	&	3.25$\pm$0.03	\\
Apr 28.93	&	44.26	&\checkmark	&	3.85$\pm$0.03	&	3.71$\pm$0.01	&	3.40$\pm$0.01	\\
May 2.93	&	48.26	&\checkmark	&	4.59$\pm$0.07	&	4.80$\pm$0.12	&	4.66$\pm$0.15	\\
May 7.99	&	53.32	&\checkmark	&	----    	&	----    	&	----    	\\	
May 8.85	&	54.18	&		&	4.04$\pm$0.11	&	3.74$\pm$0.04	&	3.34$\pm$0.07	\\
May 18.99	&	64.32	&\checkmark	&	4.40$\pm$0.16	&	4.85$\pm$0.04	&	5.15$\pm$0.04	\\
May 19.97	&	65.30	&\checkmark	&	4.93$\pm$0.02	&	5.27$\pm$0.13	&	4.93$\pm$0.10	\\
Jun 3.95	&	80.28	&\checkmark	&	4.71$\pm$0.07	&	4.04$\pm$0.01	&	3.46$\pm$0.09	\\
Jun 9.95	&	86.28	&\checkmark	&	5.49$\pm$0.07	&	4.61$\pm$0.06	&	2.66$\pm$0.05	\\
Jun 30.95       &	107.28	&		&	7.93$\pm$0.15   &       4.90$\pm$0.15   &       2.65$\pm$0.15	 \\

\hline
\end{tabular}
\label{table_ObsPhot}

\begin{list}{}{}
 \item a : Spec = Spectroscopy. Time of outburst is taken as March 15.634 UT (JD 2457097.13400).
\end{list}
\end{table}

\section{Results}
\subsection{General characteristics of the near-IR spectra}
To prevent crowding, selected  $J$, $H$ and $K$ band spectra of V5668 Sgr  are presented in Fig. 2. In addition to the strong lines due to HI, OI, CI and NI  marked on Fig. 2, weaker features may also be seen when the spectra are magnified. A detailed list of the lines is not presented here but they are typical of the FeII class of novae  (Williams, 1992) and  can be found in Das et al (2009) and Srivastava et al. (2015). In addition to lines of H, N and O, the FeII type  novae show strong lines of carbon e.g. in the $J$ band in the 1.16 to 1.18 $\mu$m wavelength region and in the $H$ band in the spectral region around Br 10 at 1.73 $\mu$ms and further redwards. These C lines are strongly seen in the present spectra. In contrast, such carbon lines are absent or very weak  in the spectra of He/N novae e.g. in the novae V597 Pup and V2491 Cyg (Naik et al. 2009). A detailed paper showing the IR classification of novae spectra and the differences between FeII and He/N novae is given in Banerjee and Ashok (2012). The FeII nova classification, independently arrived at using IR diagnostics,  is consistent with  its optical classification as a FeII type nova.

 The occurence of lines of NaI  at 2.2056
 and 2.2084 $\mu$m  are worth noting (Fig. 5) as their presence generally portends  that dust will form in a nova (Banerjee et al. 2015d). It has been shown earlier  that the  lines of NaI particularly  are associated with low excitation conditions (Das et al. 2008) implying the  existence  of a cool zone which  is conducive for dust formation. Such regions could be  associated with clumpiness in the nova ejecta which likely provide the cool dense sites needed for molecule and dust formation. Whenever these lines are seen, dust did indeed form in the nova (examples are V2274 Cyg (Rudy et al. 2003),   NQ Vul  (Ferland et al. 1979), V705 Cas (Evans et al 1996), V842 Cen (Wichmann et al. 1991), V1280 Sco (Das et al. 2008),  V5579 Sgr, V496 Sct and V5884 Sgr(Raj et al. 2012, 2014)). The  dust formation witnessed in this nova is consistent with this scenario.

\subsubsection{Evidence for a bipolar flow}
  Most of the lines showed prominent P Cygni profiles during the early stages.  However, when the lines began to be more dominantly in emission it became apparent that the profiles had, apart from a main central component, two satellite blue and red components in the wings. An example is shown in Fig. 3 for the Br$\gamma$ 2.1656 line of 2015 May 18.99 which is shown fitted by three gaussians.  The satellite components are  a strong indication of a bipolar flow likely arising from  an hour-glass morphology. Such an assumed morphology, with a dense equatorial torus and less denser bipolar lobes,  would help explain the simultaneous co-existence of dust and X-ray emission from the object as discussed in more detail in section 4.2. Similar shaped profiles were seen for example in RS Oph whose bipolar morphology was established from HST imaging (Banerjee et al. 2009 and references therein). In Fig. 3, the blue and red components are separated from the central component by -1110 and 1200 km/s respectively. The central component, which we consider to represent the principal ejecta, has an FWHM of 1060 km/s and we adopt the expansion velocity, for future calculations,  to be half the FWHM  or 530 km/s.

\subsection{Dust formation and dust mass estimate}
\label{subsec_Ncep_Lightcurve}
 Commencement of   dust grain nucleation  is first clearly seen in our data on June 9.95 wherein the $J-K$ color has increased to $\sim$ 2.85 indicating the buildup of an  IR excess.  By June 30.95, the SED is totally dominated by IR emission from dust (right panel of Fig. 4) whose  SED  is well approximated by a blackbody of 850K consistent with the SOFIA  estimate of 750K for 6 July (Gehrz et al. 2015a). In contrast to the dust SED, the SED during optical maximum (left panel, Fig. 4) shows that the radiation is  well approximated by a 9000K  blackbody  peaking in the near UV/optical.
Following Woodward et al (1993),  the mass of the dust  can be found from
$M_{dust}$ = 1.1 $\times$ 10${^{\rm 6}}$ $(\lambda F_{\lambda})_{max}$ $d^2$ / $T_{dust}^6$ .
    Here the mass of the dust shell $M_{dust}$  is in units of $M_\odot $ , $(\lambda F_{\lambda})_{max}$ is in W cm$^{-2}$  measured
at the peak of the SED, the grain temperature  $T_{dust}$ is in units of 10${^{\rm 3}}$ K, and the distance to the nova $d$ is in kpc. The formulation is valid for  dust that is composed of carbon particles of size less than 1 $\mu$m with a density of 2.25 gm cm$^{-3}$. The occurrence of strong carbon lines in the observed spectra supports our assumption that the dust could be  made up of carbon/graphite. We obtain $M_{dust}$ = 2.7 $\times$ 10${^{\rm -7}}$  $M_\odot $  for 2015 June 30.95  taking the observed parameters of
$(\lambda F_{\lambda})_{max}$ = 2.3 $\times$ 10${^{\rm -14}}$  W cm$^{-2}$, $T_{dust}$= 0.85$\times$ 10${^{\rm 3}}$
 K and $d$ = 2 kpc. Taking a canonical range  of 100 to 200 for the gas to dust ratio, we get 2.7--5.4 $\times$ 10${^{\rm -5}}$  $M_\odot $ for the mass of the gaseous component of the ejecta. This value is consistent with the typically observed range of the ejecta mass of  10${^{\rm -4}}$ to 10${^{\rm -5}}$ $M_\odot$  (e.g. della Valle et al. 2002).

 From Fig. 4 we calculate  the outburst luminosity ($L_{opt}$) and IR dust luminosity ($L_{IR}$) to be 7.2x$10{^{\rm 4}}$ and 3.71x$10{^{\rm 4}}$ $L_\odot $  respectively using a distance $d$ = 2 kpc and the relation that the outburst luminosity equals 4.03x$10{^{\rm 17}}$$d^{2}$ $ (\lambda F_{\lambda})_{max}$ (Gehrz 2008). The outburst luminosity of 7.2x$10{^{\rm 4}}$ $L_\odot $ for V5668 Sgr is roughly 1.1 times the Eddington limit for a one solar mass WD.  The ratio ( $L_{IR}$ / $L_{Opt}$)  equals  0.52 which is fairly high  although not as high as the value of  unity found in  novae like NQ Vul (Ney $\&$ Hatfield 1978) and LW Ser (Gehrz 1988) that formed optically thick dust shells. The fact that  ($L_{IR}$ / $L_{Opt}$) is relatively large  is suggestive that a large fraction of the optical radiation is intercepted by the dust and reprocessed and emitted in the IR.  However, some of the radiation does escape either because the shell is clumpy or has a partially open geometry such that it  does not fill the entire 4 $\pi$ solid angle as seen from the nova. We think there is a contribution from geometrical effects viz. the ejecta shell possibly has  a bipolar morphology
 where most of the dust emission is from a broad and dense,  optically-thick equatorial torus while the polar lobes are optically thin. The polar lobes must be the site of the X-ray emission otherwise it becomes difficult to  explain how X-rays (Page et al. 2015) and dust emission can be simultaneously co-existent ( as also seen in V339 Del which likely has a bipolar shape; Gehrz et al. 2015b, ApJ in press).  Supporting evidence for an hour-glass or bipolar geometry was seen from the shape of the line profiles as discussed earlier.

 From the right panel of Fig 4, the blackbody angular diameter $\theta_{bb}$  may be  calculated   from $\theta_{bb}$ = 2.0x$10{^{\rm 11}}$$ (\lambda F_{\lambda})_{max}^{1/2}$ $T_{bb}^{-2}$ (Ney $\&$ Hatfield, 1978) where $ (\lambda F_{\lambda})_{max}$ which equals 2.3x$10{^{\rm -14}}$ is in W cm$^{-2}$, $T_{bb}$ in Kelvin and $\theta_{bb}$ in arcseconds.  We obtain $\theta_{bb}$  = 42  milliarcsec which  used in conjunction with  an expansion velocity of 530 km/s for the ejecta   and a travel time of 107.28d for June 30.95, yields  a distance of 1.54 kpc to the nova.  This is  in good agreement  with the MMRD estimate using the work of della Valle and Livio (1995).

\begin{figure*}

\centering
\includegraphics[angle=-90,width=0.95\textwidth]{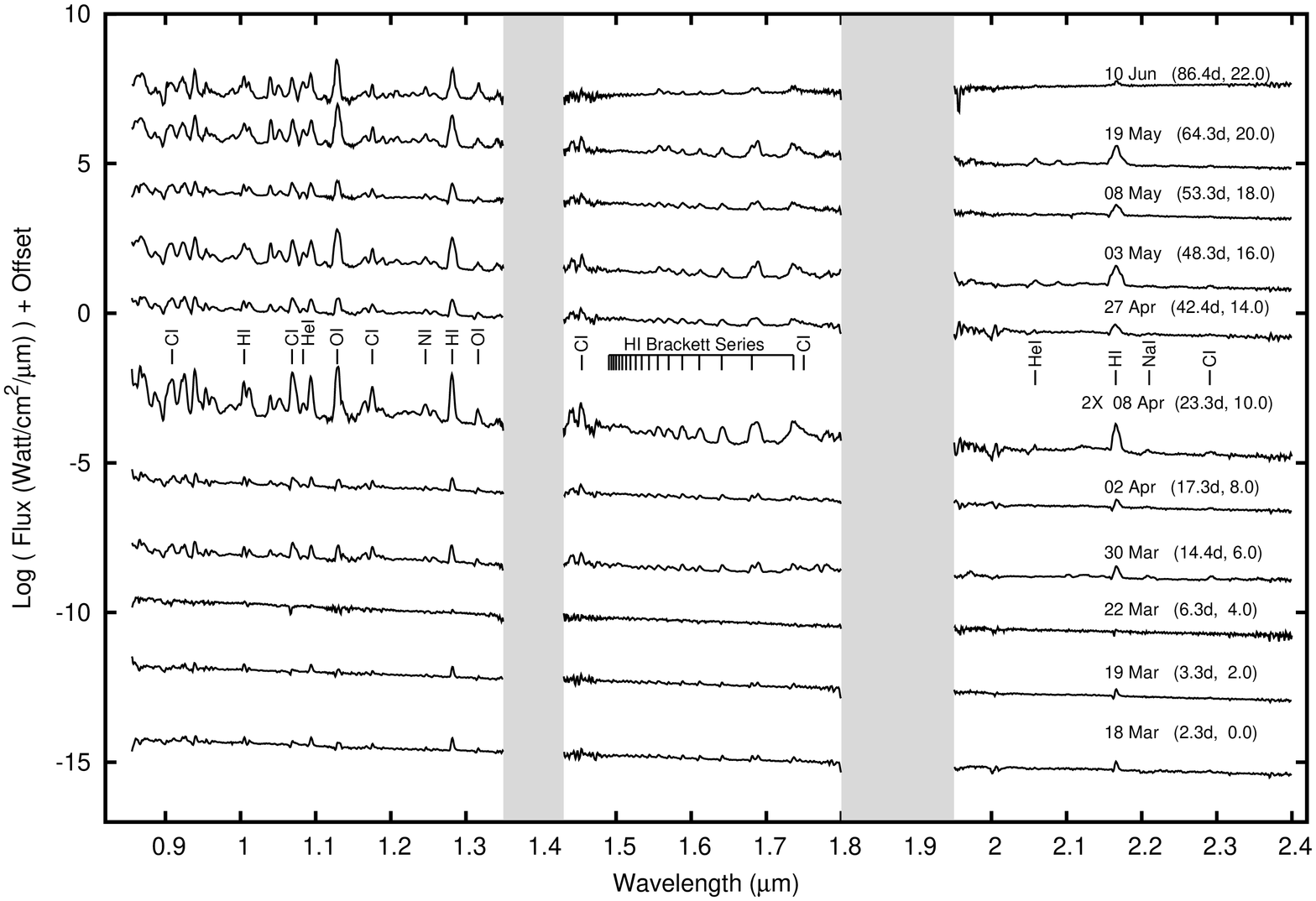}
\vskip 2mm
\caption[]{ Selected spectra of V5668 Sgr. The days after outburst and the offset are indicated in parantheses. Regions of poor atmospheric transmission are blanked out.  }
\label{PaBeta_TimeEvol}

\end{figure*}


\begin{figure}
\centering
\includegraphics[bb= 0 25 255 149, width=3.2in,clip]{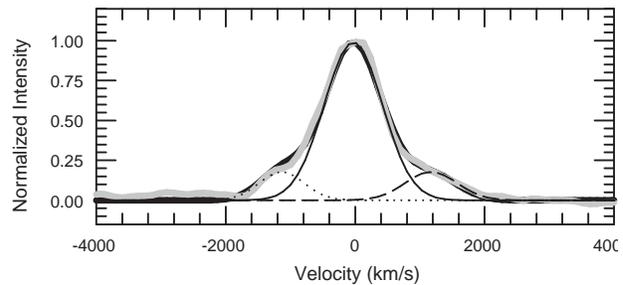}
\caption[]{A three-gaussian fit to the Br$\gamma$ 2.1656 $\mu$m profile of 2015 May 18.9 showing the
presence of satellite blue and red components on the wings suggesting  a bipolar flow. The individual gaussians(dashed, dotted $\&$ continuous lines) are shown and their coadded sum is shown by the bold black line; observed data is the bold  grey line.}
\label{}
\end{figure}


\begin{figure}
\centering
\includegraphics[bb= 0 10 313 301, width=2.75in,height=2.75in,clip]{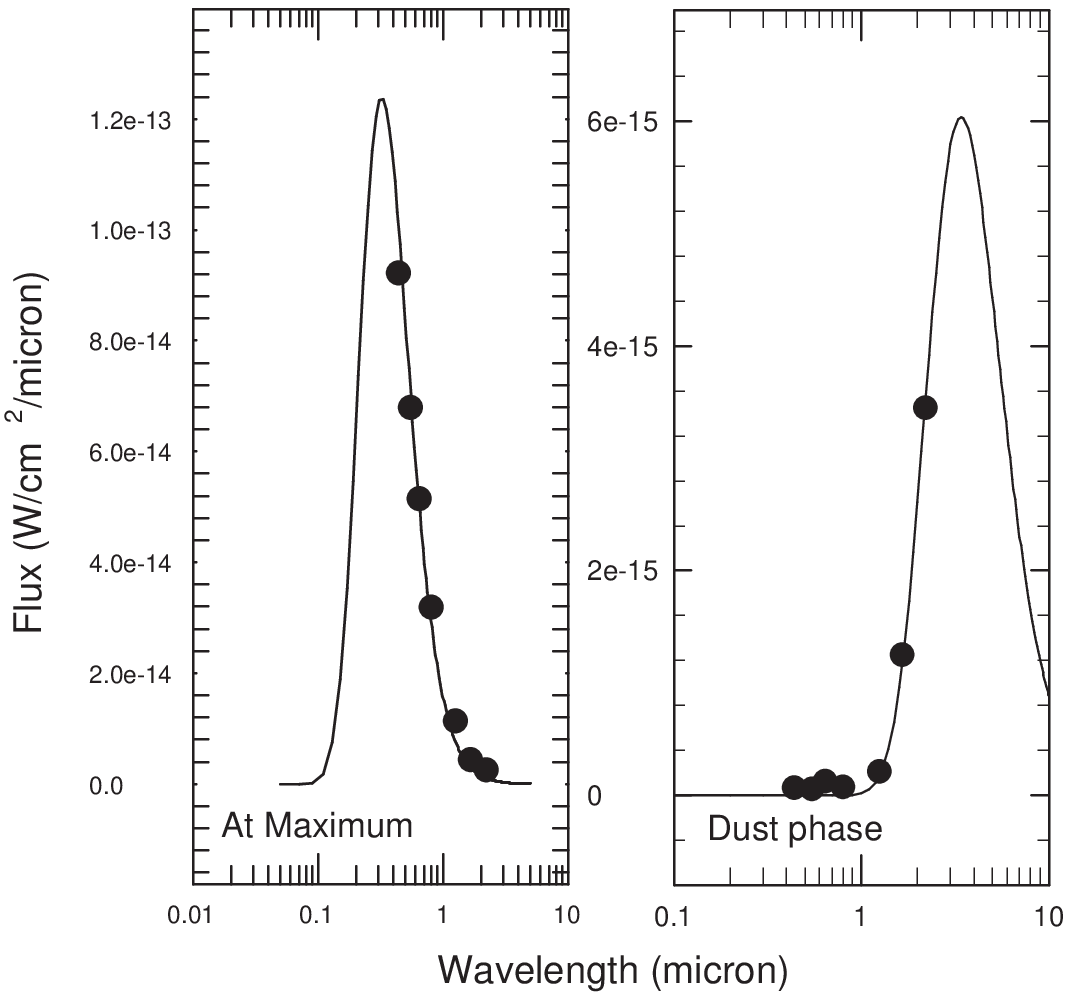}
\caption[]{ Left panel shows the SED at V band maximum using BVRI data from AAVSO and $JHK$ data from this work. A blackbody fit at 9000K is shown (continuous line). The right panel shows a similar plot during the dusty phase for the data of 30 June 2015. The SED, fitted by a blackbody at 850K, is seen to have  has shifted to the mid-IR.  }
\label{}
\end{figure}


\subsection{First overtone CO  detection and modeling}
One of the more interesting results is the detection of first overtone CO emission (Banerjee et al. 2015c) in V5668 Sgr. The most comprehensive modeling of the formation and evolution of CO in novae winds is   by Pontefract and Rawlings (2004, hereafter PR2004) who show that CO  formation should begin early after outburst and the CO abundance should remain approximately constant for 12 - 15 days after that and then it should decline fairly rapidly with time. This is true for CO formation in either an  O rich or C rich environment.

In our data CO is first seen on 2015 March 28 and is found to persist at detectable levels upto 2015 March 31 (Fig 5). A strong hint is there that CO was in emission as early as March 26  but the spectrum of this day was taken in a bright pre-dawn sky and is of unusable S/N. The duration of the CO emission thus appears to be of  at least 6 days. This duration and the rapid decline in strength are reasonably consistent with the PR2004 model timescales. The  emission is weaker than in V2615 Oph or V496 Sct  although in Banerjee et al. (2015c) we mistakenly reported the detection as strong because of ratioing the nova spectrum with the  wrong standard star spectrum. For modeling the emission parameters, we used the same LTE  model developed and discussed in great details in  Das et al (2009) for V2615 Oph and subsequently applied to V496 Sct and V5584 Sct. Compared to V2615 Oph,  the S/N of the spectra is poor here due to the  low elevation at which the source was studied, and thus not conducive for accurate modeling.   The data  essentially covers only  three of the bands ($\nu$ = 2-0, 3-1, 4-2) and additional  C I lines at 2.2906 and 2.3130 $\mu$m further  complicate the modeling. The strong 2.2906 $\mu$m CI line causes the poor fit of the 2-0 band (bandhead at 2.2935 $\mu$m)   in Fig. 5 wherein we show observed and model fits for 28, 29 and 30 March. Allowing for all these factors, from the model fits, we estimate   the mass of the CO gas $M_{CO}$ on 28, 29 and 30 March to be (2.0, 1.0 and 0.5) $\times$ 10$^{-8}$($d/2kpc$)$^{2}$ M$_\odot$ respectively where $d$ = 2kpc; the temperature on all three days is estimated to be of 4000 $\pm$ 300 K. Knowing $M_{CO}$, the CO column density is calculated assuming  that the CO is in a  shell of radius R which is  estimated  knowing the time  since outburst and the velocity of the shell. In this manner we determine the CO column densities to be (1.94, 0.83 and 0.36) $\times$ 10$^{19}$ cm$^{-2}$   on the three days respectively. The $^{12}$C/$^{13}$C   ratio is difficult to constrain accurately because only the 2-0 band of $^{13}$CO at 2.3448$\mu$m is effectively available for setting a constraint. Our model fits indicate   a $^{12}$C/$^{13}$C ratio of $\sim$ 1.5;  other known estimates
 of the $^{12}$C/$^{13}$C  ratio are listed in Table 2.

\begin{table}
\caption[]{Galactic Novae which have shown first-overtone CO emission.}
\begin{tabular}{lccl}
\hline\\
Nova & Detection$^a$  &  $^{12}$C/$^{13}$C& Reference \\
&  epoch(d)& &\\
\hline \\

NQ Vul               &         19&$\geq$3    & Ferland 1979\\
V842 Cen          &         25&$\sim$2.9    &  Wichmann 1991\\
V705 Cas          &           6&$\geq5$    &  Evans 1996  \\
V2274 Cyg        &         17&$\sim$1.2    & Rudy 2003          \\
V2615 Oph        &          9&$\geq$2   & Das 2009\\
V5584 Sgr         &        12  &   & Raj 2014\\
V496 Sct           &        19&$\geq$1.5    & Raj 2012; Rudy 2009\\
V2676 Oph       &        37     & & Rudy 2012a    \\
V1724 Aql       &          7      &&  Rudy 2012b \\
V5668 Sgr       &          12    & $\sim$1.5& this work \\
\hline
\end{tabular}
\label{table4}
\begin{list}{}{}
 \item a : In terms of days after outburst.
\end{list}
\end{table}

\begin{figure}
\centering
\includegraphics[bb= 0 0 310 282, width=3.0in,clip]{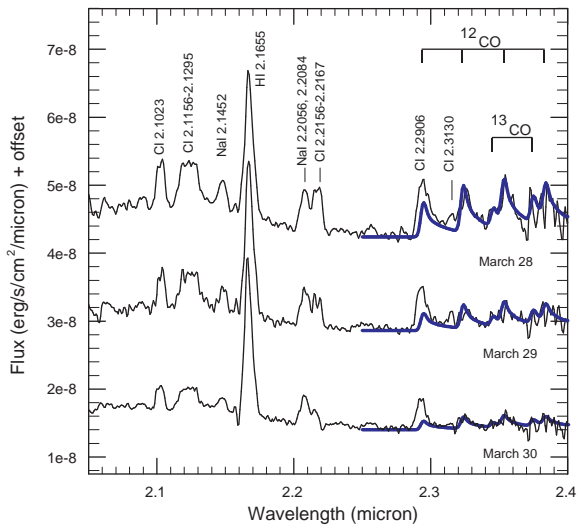}
\caption[]{Observed and modeled CO emission (see section 4.3 for details)}
\label{fig_NSco_FWHMvsTime}
\end{figure}

\section{Summary}
 We present multi-epoch NIR observations of Nova V5668 Sgr. The spectra indicate it to be a FeII class of nova.  The carbon monoxide first overtone bands are modeled  to estimate the CO mass, temperature, column density and  $^{12}$C/$^{13}$C ratio.  V5668 Sgr was a strong dust producer. Analysis of the dust SED is made to estimate the dust mass,  blackbody angular diameter of the dust shell
and the distance to the nova.


\section{Acknowledgments}
 The research work at the PRL is funded by the Dept. of Space, Government of India. We gratefully acknowledge the use of AAVSO data.  

\end{document}